# Effect of Rashba spin-orbit coupling on Faraday rotation in an extended Haldane model*

Yuan Fang(方圆)[1, 2,†], Yixiang Wang(王义翔)[3], Xiaopu Zhang(张孝谱)[1, 2,‡]

1. Laboratory of Thin Film Optics, Key Laboratory of Materials for High Power Laser, Shanghai Institute of Optics and Fine Mechanics, Shanghai 201800, China

2. Key Laboratory of Materials for High Power Laser, Chinese Academy of Sciences, Shanghai 201800, China

3. School of Science, Jiangnan University, Wuxi 214122, China

Utilization of Faraday rotation (FR) properties of topological materials offers a promising route toward novel magneto-optical devices. We systematically investigated the effect of Rashba spin-orbit coupling (SOC) on FR spectra in an extended Haldane model, which incorporates Rashba SOC and exchange splitting into the original spinless Haldane framework. Using the Kubo formalism, we calculated the FR spectra across the model's rich topological phase diagram. We found that in the Chern number C=2 region, in the absence of exchange splitting, the FR angle can exceed 4° and its peak position is tunable by the Rashba SOC. In contrast, with the inclusion of exchange splitting, a nearly flat FR profile emerges over a broad frequency range, and the FR peak values increase monotonically with the Rashba SOC strength. The Rashba SOC opens additional transition channels, whose net contribution constructively enhances the FR peak. Furthermore, we derived a low-energy effective Hamiltonian expanded up to quadratic terms, the results of which are in good agreement with tight-binding model calculations, thereby validating our numerical results. Our findings suggest that magneto-optical device characteristics can be designed and optimized through Rashba SOC engineering.

* Project supported by the National Natural Science Foundation of China (Grant No. 52573361).

† Corresponding author. E-mail: yfangphysics@zju.edu.cn

‡ Corresponding author. E-mail: zhangxiaopu@siom.ac.cn

## 1. Introduction

In magneto-optics, Faraday rotation (FR) refers to the rotation of the plane of polarization of incident linearly polarized light (which is a sum of left and right-handed circularly polarized lights) when it traverses a transparent material[1]. In addition to this rotation, the transmitted polarization acquires a certain degree of ellipticity. With the rapid advancement of two-dimensional (2D) materials science, FR in graphene and graphene-like materials has garnered considerable research interest due to its potential applications in passive and active magneto-optical devices[2–6]. Experimentally, a maximum FR angle exceeding ~6° has been observed from a single graphene layer under a 7 T magnetic field[2]. Near optimal non-reciprocal isolator devices have been successfully demonstrated using three graphene sheets separated by thin poly methyl methacrylate (PMMA) layers[4]. From a theoretical perspective, silicene in the quantum spin Hall insulator state is predicted to have a large FR of 4.5° at 5 T[6] .

Spin-orbit coupling (SOC) has long been recognized as a crucial factor influencing the FR properties of materials[1,7] . However, the precise role of SOC in FR remains a subject of debate. For instance, bismuth-substituted yttrium iron garnet (Bi:YIG) exhibits FR enhancement attributed to increased SOC[8,9] . Theoretical investigations of transition metal systems suggest that magneto-optical behavior depends linearly on the intrinsic SOC strength and is dominated by spin-conserving processes[10]. More recently, it was theoretically shown that the FR angle of monolayer $MoS_2$ at terahertz frequencies can be enhanced and controlled by applying Rashba SOC and magnetic field, an effect arising from spin-flip optical transitions[11] . Conversely, quantitative calculations for praseodymium-substituted yttrium iron garnet (Pr:YIG) and $PrF_3$ indicate that FR does not scale linearly with SOC strength, and that SOC is not strictly necessary to generate FR[12,13] . Moreover, topological magneto-optical effects have been demonstrated in noncoplanar antiferromagnets due to finite scalar spin chirality, without any reference to SOC or exchange splitting[14] .

Recent studies have revealed that the Haldane model, when extended to include Rashba SOC and exchange splitting, can host rich topological phases characterized by Chern numbers up to 4[15,16] . The Haldane model on a honeycomb lattice serves as a paradigmatic framework for realizing topologically distinct phases of matter[17] and provides the conceptual foundation for exploring topological insulators and superconductors[18,19]. Rashba SOC, a distinct type of SOC, induces momentum-dependent spin splitting and can be generated by an external electric field or adatoms, without requiring low-temperature operation[20,21]. The question of how the FR properties of the Haldane model are influenced and potentially controlled by Rashba SOC is both intriguing and, to the best of our knowledge, remains largely unexplored.

In this paper, we systematically investigated the effect of Rashba SOC on the FR properties of an extended Haldane model using the Kubo formalism. We calculated FR spectra at selected points in the phase diagrams corresponding to different Rashba SOC strengths, with and without exchange splitting or on-site energy differences. We identified particularly interesting FR behaviors in the Chern number C=2 region. In the absence of exchange splitting, the FR angle can exceed 4°, and its peak position is modulated by the Rashba SOC strength. With exchange splitting, the FR spectra exhibit a nearly flat region over a broad frequency range, and the FR peak values can be monotonically enhanced by increasing Rashba SOC. By decomposing the optical Hall conductivity into different spin channels, we showed that pure spin-conserving, pure spin-mixing and mixed spin-conserving/spin-flip channels all enhance the FR peak, whereas the pure spin-flip process yields an opposite contribution. We also derived the low-energy effective Hamiltonian of the system and demonstrated that it matches reasonably well with the tight-binding model when expanded up to the quadratic terms in momentum.

## 2. Model and Methods

We adopt the following tight-binding Hamiltonian which incorporates Rashba SOC and exchange splitting into the original spinless Haldane model[5,16,22] :

$$H = -t_1 \sum_{<ij>\alpha} c_{i\alpha}^{+} c_{j\alpha} + i\lambda_R \sum_{<ij>\alpha\beta} \left(\boldsymbol{\sigma} \times \widehat{\boldsymbol{d}_{ij}}\right)_{\alpha\beta}^{z} c_{i\alpha}^{+} c_{j\beta}$$

$$+t_2 \sum_{\ll ij \gg \alpha} e^{i\nu_{ij}\varphi} c_{i\alpha}^{+} c_{j\alpha} + M \sum_{i\alpha} \tau_i c_{i\alpha}^{+} c_{i\alpha} + \lambda_{FM} \sum_{i\alpha} c_{i\alpha}^{+} (\sigma_z)_{\alpha\alpha} c_{i\alpha}, \qquad (1)$$

where $c_{i\alpha}^{+}$ $(c_{i\alpha})$ is the creation (annihilation) operator for site i and spin $\alpha$. $< i,j >$ $(\ll i,j \gg)$ represents the summation over all the nearest-neighbor (next-nearest-neighbor) sites. $\boldsymbol{\sigma} = (\sigma_x,\ \sigma_y,\ \sigma_z)$ are the Pauli matrices for spin. $\widehat{\boldsymbol{d}_{ij}} = \boldsymbol{d}_{ij}/|\boldsymbol{d}_{ij}|$ is the unit vector from sites i to j. The first term in the Hamiltonian denotes the nearest-neighbor (NN) interactions with the hopping integral $t_1$ >0. The second term corresponds to the Rashba SOC. The third term describes the next-nearest-neighbor (NNN) interactions with the hopping integral $t_2$, where $\nu_{ij} = +1\ (-1)$ for clockwise (anticlockwise) NNN hopping direction, and we set $\varphi = \frac{\pi}{2}$. In the fourth term, $\tau_i = +1\ (-1)$ for the A (B) sublattice, representing an on-site energy difference for the A (+M) and B (-M) sublattices. The fifth term introduces an exchange splitting strength $\lambda_{FM}$, which can be generated by doping the system with magnetic ions. It is important to note that only internal magnetization is considered here, and no external magnetic field is applied; consequently, Landau levels are not considered in this work. In the following, we set $t_1$=1 as the energy unit, normalizing all other parameters accordingly. The Fermi level is consistently set to zero energy.

For the noninteracting multiband 2D system, ignoring intraband transitions, the conductivity tensor is given by Kubo-Greenwood formula[5,23] :

$$\sigma_{\mu\nu}(\omega) = \frac{e^2\hbar}{iS} \sum_{\boldsymbol{k}} \sum_{mn} \frac{n_F\left(\epsilon_n(\boldsymbol{k})\right) - n_F\left(\epsilon_m(\boldsymbol{k})\right)}{\epsilon_n(\boldsymbol{k}) - \epsilon_m(\boldsymbol{k})} \times \frac{\left[v_\mu(\boldsymbol{k})\right]_{nm} \left[v_\nu(\boldsymbol{k})\right]_{mn}}{\epsilon_n(\boldsymbol{k}) - \epsilon_m(\boldsymbol{k}) + \hbar\omega + i\eta}, \qquad (2)$$

where $\epsilon_n(\boldsymbol{k})$ denotes the dispersion of the nth band, and the $\boldsymbol{k}$ summation runs over the first Brillouin zone. $n_F(\epsilon)$ is the Fermi-Dirac distribution function. The velocity operator in the direction of $\mu$, $v_\mu$ $(\mu = x,\ y)$ is defined by

$$v_\mu \equiv \frac{1}{\hbar} \frac{\partial H}{\partial k_\mu}, \qquad (3)$$

The small number $\eta$ represents the inelastic scattering rate or quasiparticle lifetime which we take as $0.01t_1$ in all the following calculations, consistent with experimental results for graphene[24] . S is the area of the 2D material.

The Berry curvature of the filled bands is defined as:

$$\Omega(\boldsymbol{k}) = i\hbar^2 \sum_{mn} \frac{n_F(\epsilon_n(\boldsymbol{k})) - n_F(\epsilon_m(\boldsymbol{k}))}{\epsilon_n(\boldsymbol{k}) - \epsilon_m(\boldsymbol{k})} \times \frac{[v_\mu(\boldsymbol{k})]_{nm}[v_\nu(\boldsymbol{k})]_{mn}}{\epsilon_n(\boldsymbol{k}) - \epsilon_m(\boldsymbol{k})}, \tag{4}$$

The Chern number is used to characterize the topological phases. It is calculated from the zero frequency Hall conductivity $\sigma_{xy}$ using the relation:

$$-\frac{e^2}{h} C = \sigma_{xy}(0), \tag{5}$$

Because the closing and reopening of the bulk band gap signifies a topological phase transition, we present diagrams of the bulk band gap in units of eV in the $(\lambda_R, t_2)$ plane, as shown in Figure 1-4(a). Different topological regions are subsequently labelled by their corresponding Chern numbers.

Using the thin film approximation[2,24] , the FR angle $\theta_F$ can be expressed as:

$$\theta_F = \frac{1}{2}(\theta_+ - \theta_-), \tag{6}$$

where

$$\theta_\pm = -arctan \frac{\mu_0 c \sigma''_\mp}{1 + \sqrt{\epsilon_r} + c\mu_0 \sigma'_\mp}, \tag{7}$$

and $\sigma_\pm = \sigma'_\pm + i\sigma''_\pm$. Here, $\sigma'_\pm$ and $\sigma''_\pm$ are the real and imaginary parts of $\sigma_\pm$, respectively. Explicitly, we have

$$\sigma_\pm = (\sigma'_{xx} \mp \sigma''_{xy}) + i(\sigma''_{xx} \pm \sigma'_{xy}), \tag{8}$$

Under the assumptions that $\theta_F \leq 1$ and $1 + \sqrt{\epsilon_r} \gg c\mu_0 \sigma'_\mp$ (the latter assumption is the more stringent of the two), the following approximate expression can be obtained[2,24]:

$$\theta_F \approx \frac{\mu_0 c}{1 + \sqrt{\epsilon_r}} \sigma'_{xy}, \tag{9}$$

In the above formulas, $\epsilon_r$ is the relative permittivity of the dielectric substrate on which the 2D material is deposited. For simplicity, we set $\epsilon_r = 1$ in all subsequent results[25] . The exact expression for the FR angle is used for all the tight-binding model calculations. The approximate expression is validated by comparison with the exact results as shown in Fig. S1 and is employed for the low-energy effective model calculations presented later.

## 3. Results and Discussions

### 3.1 Topological phase diagrams and Faraday rotation spectra

By combining the Chern number and bulk band gap, we first presented the phase diagrams for the extended Haldane model with respect to Rashba SOC and NNN hopping integral in Fig. 1(a) for $\lambda_{FM} = 0$ and $M = 0$, i.e., in the absence of exchange splitting and on-site energy difference. Within this parameter range, only a single topological phase with Chern number $C = 2$ is observed. We selected four representative points with $t_2$=0.15 and $\lambda_R = 0, 0.2, 0.4, 0.6$, indicated by red dots in the phase diagram, to calculate the Faraday rotation spectra shown in Fig 1(b). Note that due to symmetry with respect to the sign of $\lambda_R$, in the following we only considered these four positive $\lambda_R$ values unless otherwise stated. Owing to the nonzero Chern number, the Faraday rotation angle $\theta_F$ is non-zero even at zero frequency. A clear trend in the spectra evolution is that the FR angle peak positions shift toward lower photon energies with increasing Rashba SOC strength. This behavior correlates with the decrease in the bulk band gap as Rashba SOC increases. Similar SOC-induced shifts in peak positions have been observed in the optical spin Hall conductivity of Kagome lattices[26] . This is intuitively understood as follows: when the photon energy approaches the band gap value, the denominator in the Kubo-Greenwood formula induces a resonance corresponding to magneto-optical transitions. This also suggests that the peak position can serve as a fingerprint of the Rashba SOC strength in this regime. Notably, for $\lambda_R = 0.2$, the maximum absolute FR angle exceeds 4°.

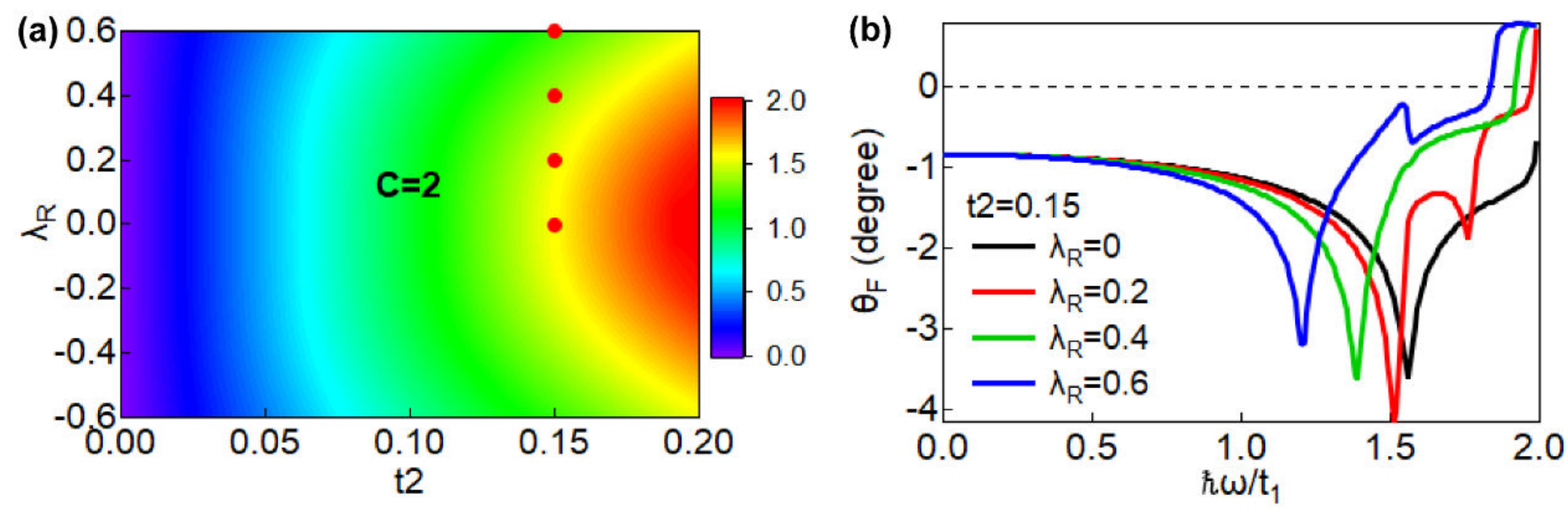

Fig. 1. (color online) Phase diagram and FR spectra for $\lambda_{FM} = 0$ and $M = 0$. (a) $\lambda_R$-$t_2$ phase diagram. Red dots indicate the parameter points where FR curves were calculated. (b) FR spectra calculated for $t_2$=0.15 and various $\lambda_R$ values.

Maintaining zero exchange splitting but introducing an on-site energy difference $M = 0.2$, we obtained the phase diagram shown in Fig. 2(a). A phase boundary with a zero band gap emerges. On the right-hand side of this boundary, the Chern number remains $C = 2$, while on the left-hand side it becomes $C = 0$. For the $C = 0$ phase, we set $t_2 = 0.015$ and presented the results in Fig 2(b). The FR angles at zero frequency are now zero but acquire finite values at non-zero frequencies. A unique feature of these spectra is the emergence of two additional peaks at finite Rashba SOC. This is induced by Rashba SOC-driven energy level splitting around $K$ and $K'$ points in the momentum space. For $\lambda_R = 0.4$, the bulk band structure plotted along $K - \Gamma - K'$ direction, shown in Fig. 2(d), exhibits the characteristic Rashba band splitting. At the $K$ and $K'$ points, transition energies of $\Delta \cong 1.51 t_1$ and $1.33 t_1$, respectively, correspond to the additional peak and valley structures observed in the FR spectra of Fig. 2(b). For the $C = 2$ region, we set $t_2 = 0.15$ and showed the results in Fig. 2(c). Compared to the previous case with $\lambda_{FM} = 0$ and $M = 0$, the peak positions again shift toward lower photon energies with increasing Rashba SOC, but now two sets of peaks are present instead of one.

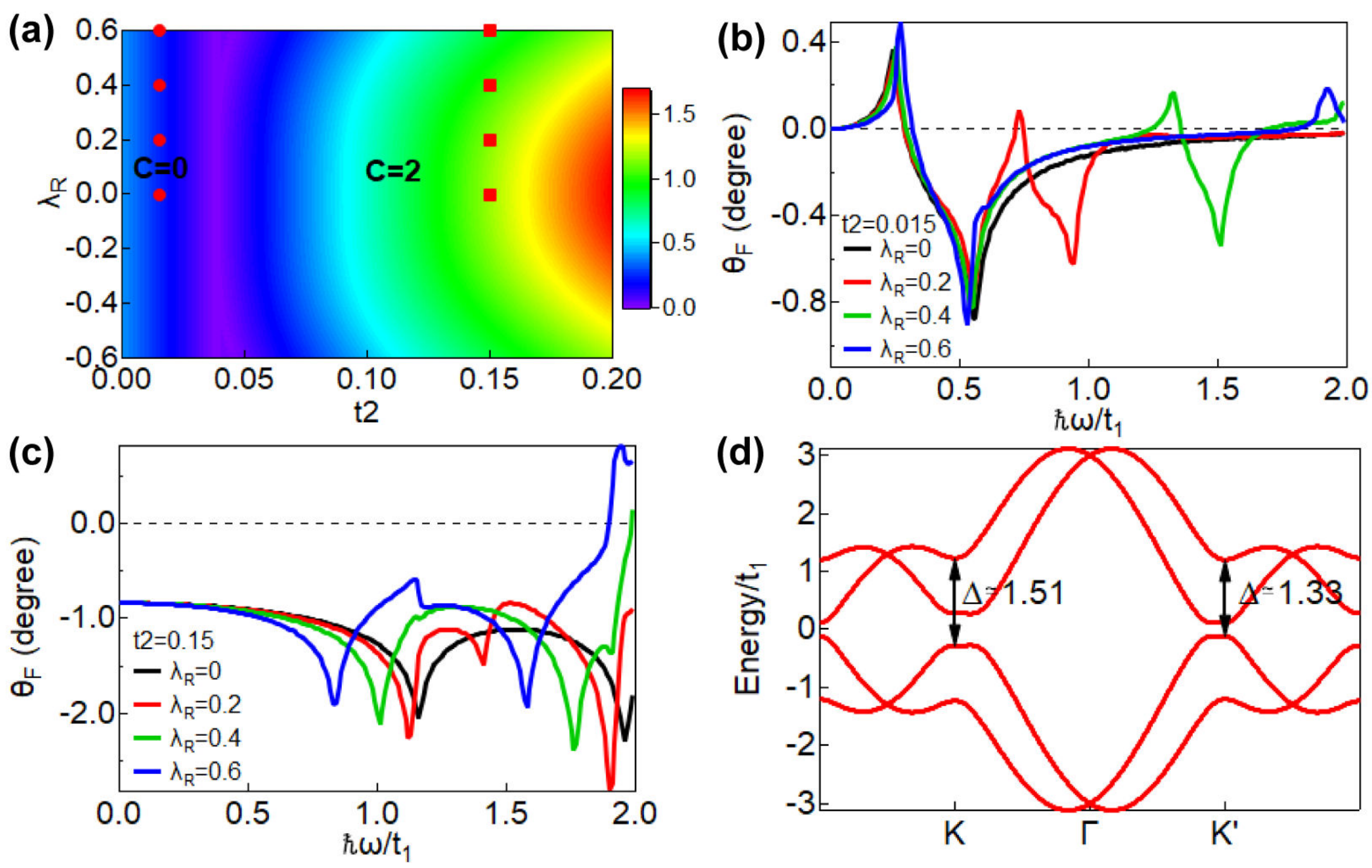

Fig. 2. (color online) Phase diagram and FR spectra for $\lambda_{FM} = 0$ and $M = 0.2$. (a) $\lambda_R$-$t_2$ phase diagram. Red dots mark the parameter points for FR calculations. FR spectra calculated for $t_2 = 0.015$ (b) and $t_2 = 0.15$ (c) with various $\lambda_R$ values. (d) Bulk band structure along the $K$-$\Gamma$-$K'$ direction for $t_2$=0.015 and $\lambda_R = 0.4$.

For the case with exchange splitting $\lambda_{FM} = -0.5$ and zero on-site energy difference $M = 0$, the phase diagram and corresponding FR angle spectra are shown in Fig. 3. Fig. 3(a) reveals a rich topological landscape with phases characterized by Chern numbers $C = -2, 2, 4$. It is natural to think that a spinless Haldane model can have Chern number up to 1 and a spinful Haldane model can have Chern number up to 2. Here the origin of the unusual $C = 4$ phase is the existence of four valley edge states propagating in the same direction with the coexistence of Rashba SOC and exchange splitting[16] . For $C = -2$ and $C = 4$ phase, $\lambda_R = 0$ is the phase boundary so we only took $\lambda_R = 0.2, 0.4, 0.6$ to calculate $\theta_F$. For $t_2$=0.015, i.e., $C = -2$ phase, as shown in Fig. 3(b) $\theta_F$ initially takes positive values and changes sign at larger frequency. For $t_2 = 0.089$, which corresponds to the largest Chern number C=4, $\theta_F$ has the large values in the terahertz region up to ~2°, as shown in Fig. 3(c). A particularly intriguing phenomenon is observed in the $C = 2$ region for $t_2 = 0.15$, as shown in Fig. 3(d). A nearly flat FR curve appears over a broad frequency range, and the FR peak values increase monotonically with the Rashba SOC strength. This behavior suggests that the FR angle can be modulated by an extrinsically controlled electric field (which tunes the Rashba SOC), potentially inspiring device applications such as optical isolators. This can be understood by noting that, for this parameter set, the exchange splitting tends to align the spin along the z-direction whereas the Rashba SOC favors in-plane spin alignment with spin-momentum locking. Their coexistence and competition induce spin mixing and spin-flip transitions, thereby activating initially forbidden optical transitions. A detailed analysis will be provided later.

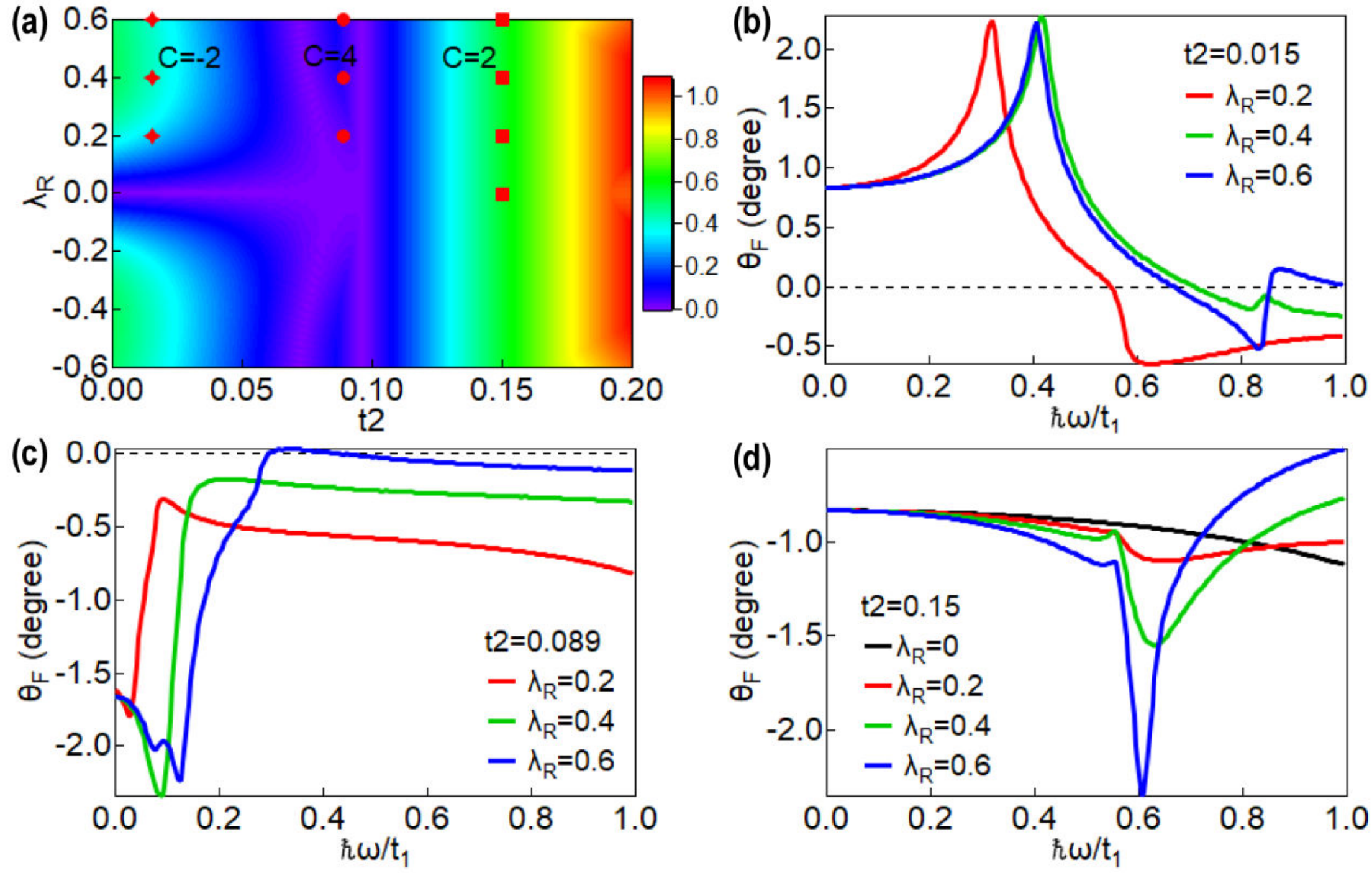

Fig. 3. (color online) Phase diagram and FR spectra for $\lambda_{FM} = -0.5$ and $M = 0$. (a) $\lambda_R$-$t_2$ phase diagram. Red dots represent the points where FR curves were calculated. FR spectra were calculated at $t_2$=0.015 (b), $t_2$=0.089 (c) and $t_2$=0.15 (d) for different $\lambda_R$ values.

For the case incorporating both exchange splitting $\lambda_{FM} = -0.5$ and an on-site energy difference $M = 0.2$, the phase diagram and corresponding FR spectra are shown in Fig. 4. As shown in Fig. 4(a), even more topological phases emerge, including those with $\mathrm{C} = -2, 0, 1, 2, 3$. The origin of the $\mathrm{C} = 3$ phase is attributed to the coexistence of two valley edge states and one non-valley edge state[16]. It is noteworthy that the phase boundary between $C = -2$ and $C = 1$ phases exhibits a finite slope, meaning that for a fixed $t_2$ value, a phase transition can be induced solely by varying $\lambda_R$. Generally, as seen in Fig. 4(b-f), both the FR angle magnitudes and the first set peak positions are reduced compared to the case without an on-site energy difference. The FR spectra in Fig. 4(f) resemble those in Fig. 3(d), as they both feature a partially flat region and peak values that monotonically increase with Rashba SOC strength, although the flat frequency range is narrower and the maximum peak values are smaller in this case.

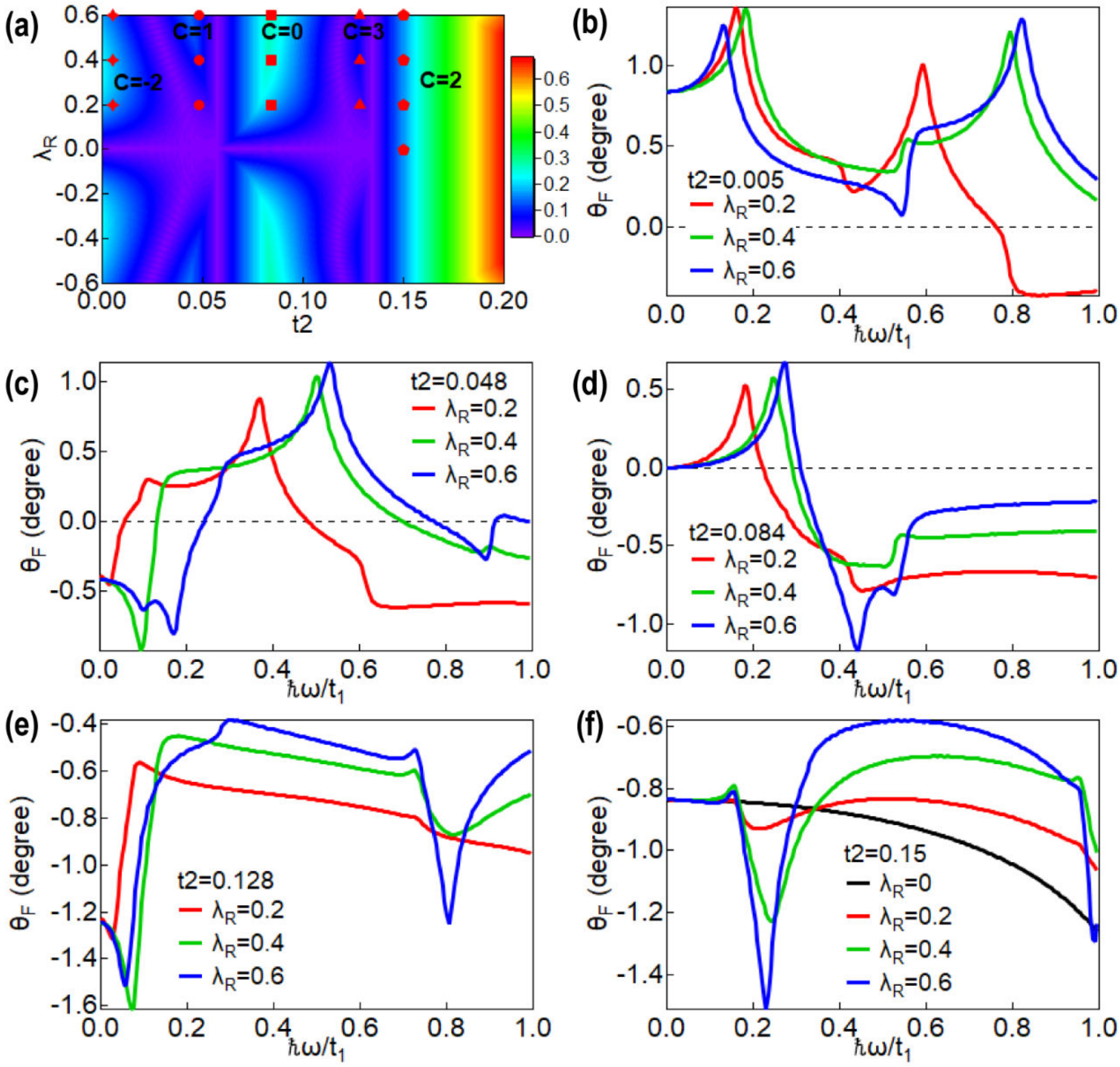

Fig. 4. (color online) Phase diagram and FR spectra for $\lambda_{FM} = -0.5$ and $M = 0.2$. (a) $\lambda_R$-$t_2$ phase diagram. Red dots indicate points for FR calculations. FR spectra were calculated at $t_2$=0.005 (b), $t_2$=0.048 (c), $t_2$=0.084 (d), $t_2$=0.128 (e) and $t_2$=0.15 (f) for different $\lambda_R$ values.

Before going to analyze the FR peak enhancement with increasing Rashba SOC strength, we want to mention that based on the results shown in Fig. S2 the FR angle peak positions correspond to the uprise turning positions in the joint density of states (JDOS) which is defined as[27]:

$$J(\omega)=\sum_{k}\sum_{\substack{n\in valence\\ m\in conduction}}\frac{1}{N_k\pi}\frac{\eta}{\left(\hbar\omega+\epsilon_n(\boldsymbol{k})-\epsilon_m(\boldsymbol{k})\right)^2+\eta^2} \tag{10}$$

where we numerically approximated the Dirac delta function using the Lorentzian function. $N_k$ is the number of k mesh points in the numerical calculation and $\eta$ is taken as $0.01t_1$. This suggests that the origin of the emergence of FR peak is closely related with the resonance condition when photon energies match van Hove singularity energy differences[28]. Similar relationship has been reported in [5].

### 3.2 Analysis of the Faraday rotation peak enhancement

To further analyze the FR peak enhancement with increasing Rashba SOC strength observed in Fig. 3(d) and Fig. 4(f), the FR peak values were plotted as a function of $\lambda_R$ in Fig 5(a) and 5(b), respectively. The FR angles were calculated at photon energy $0.605t_1$ for Fig. 5(a) and $0.23t_1$ for Fig. 5(b). The relationship between the FR peak value and Rashba SOC strength can be well fitted by parabolic curves in both cases, indicating that the enhancement follows an approximately quadratic trend. As shown in Fig. S3, when we took the square root of $-\theta_F$ minus its value at $\lambda_R=0$, the obtained curve follows linear trend. This suggests the dependence of FR peak value upon Rashba SOC studied here does not contain leading linear terms. Actually, the absence of linear term is a general property for the model studied here as proved in the supplementary material based on the low-energy effective model presented in section 3.3 and perturbation theory. We want to mention that higher order terms are not forbidden so that the FR angle relationship with respect to Rashba SOC strength at a fixed frequency is not always quadratic. To verify the generality of the peak enhancement, we calculated

FR spectra at different $t_2$ values within the $\mathrm{C}=2$ region of the phase diagrams. For $\lambda_{FM}=-0.5$ and $M=0$, we chose $t_2=0.125, 0.15$ and $0.175$. For $\lambda_{FM}=-0.5$ and $M=0.2$, we selected $t_2=0.15, 0.175$ and $0.2$. As can be seen from Fig 5, the enhancement of the FR peak with increasing Rashba SOC strength is a ubiquitous feature across these different $t_2$ values. Furthermore, the peak centers are closely aligned in energy near the $\mathrm{C}=2$ phase boundary, while they become more separated as one moves away from the phase boundary. Note that with changing parameters the FR angle peaks have an upper bound if the system has a band gap and finite $\eta$ which is proved in the supplementary material. In fact, for $\lambda_R=0.8$, the FR peak already begins to drop compared with $\lambda_R=0.6$ as shown in Fig. S4.

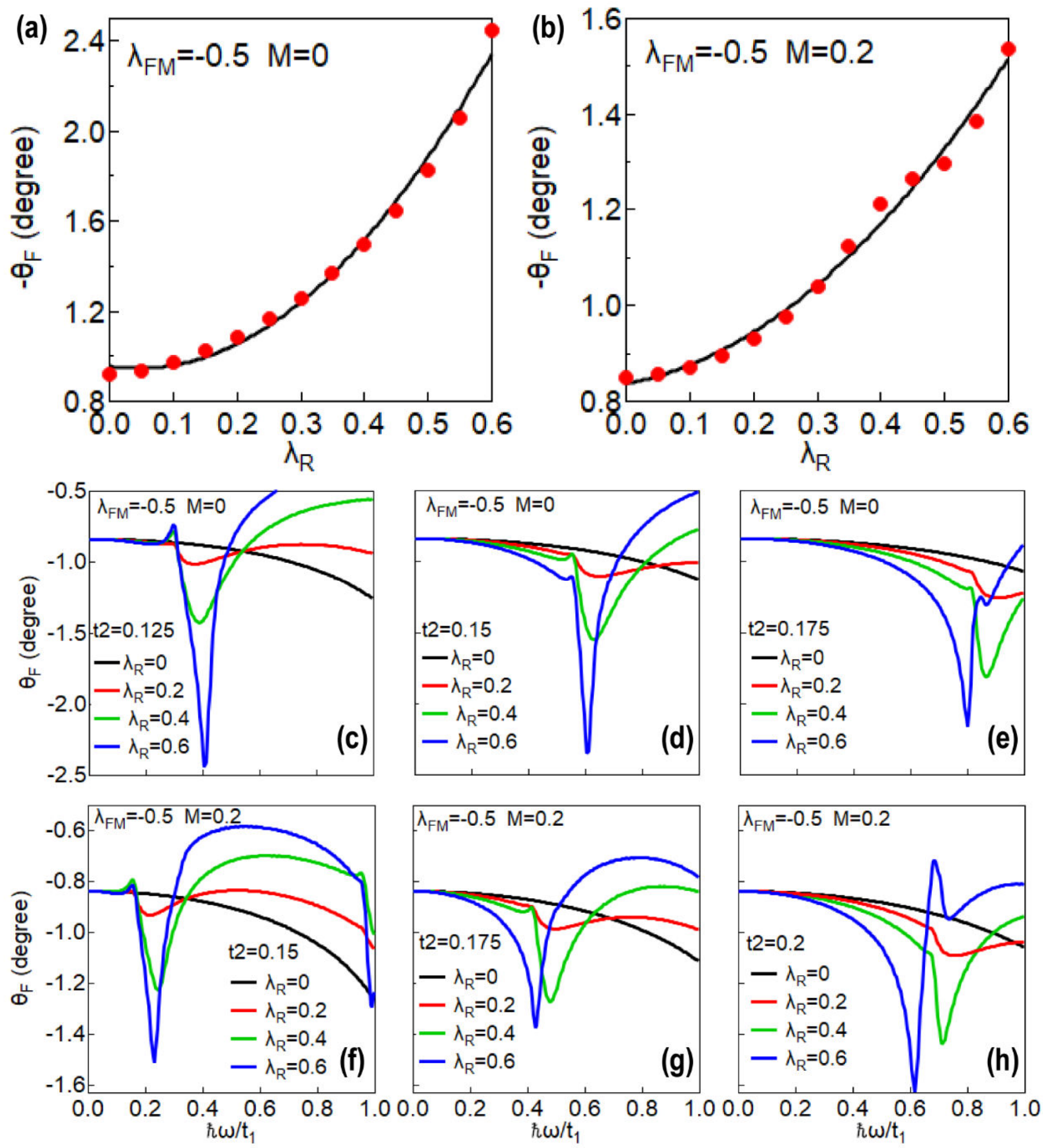

Fig. 5. (color online) The FR peak enhancement with Rashba SOC. (a) FR peak value as a function of Rashba SOC for $\lambda_{FM} = -0.5$, $M = 0$ and $t_2$=0.15. (b) FR peak value vs. Rashba SOC for $\lambda_{FM} = -0.5$, $M = 0.2$ and $t_2$=0.15. Red dots represent numerical results and black curves are parabolic fits. FR curves for $C = 2$ region with $\lambda_{FM} = -0.5$ and $M = 0$ (c-e) and $\lambda_{FM} = -0.5$ and $M = 0.2$ (f-h) for different $t_2$ values.

To elucidate the origin of the enhanced Faraday rotation peak value with respect to Rashba SOC in the $C = 2$ phase in the presence of exchange splitting, we calculated the Berry curvature and the finite-frequency optical transition matrix element at $0.605t_1$ [29,30]. As seen in Fig. 6 (a) and 6(b), for $\lambda_R = 0$, there is almost no enhancement of the optical transition matrix element at the peak frequency (~$0.605t_1$) relative to the Berry curvature. Fig. 6 (c) shows the expectation value of $S_z$ along the dashed lines in the momentum space marked in Fig. 6 (a) and 6(b). The electron spin is purely up (red) or down (blue), a direct consequence of the exchange splitting in the absence of Rashba SOC. In contrast, as can be seen from Fig. 6 (d) and (e) for $\lambda_R = 0.6$, a clear enhancement of the optical transition matrix element is observed at the peak frequency compared to the Berry curvature. In fact, the intensity at the hot spots increases by more than an order of magnitude. Calculation of the expectation value of $S_y$ along the same dashed lines passing these hot spots reveal a substantial $S_y$ component, as shown in Fig. 6(f). This behavior arises from the competition between out-of-plane spin alignment induced by exchange splitting and the in-plane spin alignment favored by Rashba SOC. Moreover, it is evident from the tight-binding Hamiltonian that the Rashba SOC term inherently includes spin-flip transitions. Consequently, originally forbidden transition channels are opened due to the Rashba SOC-induced spin mixing and spin-flip transitions, leading to the enhancement of the Faraday rotation peak.

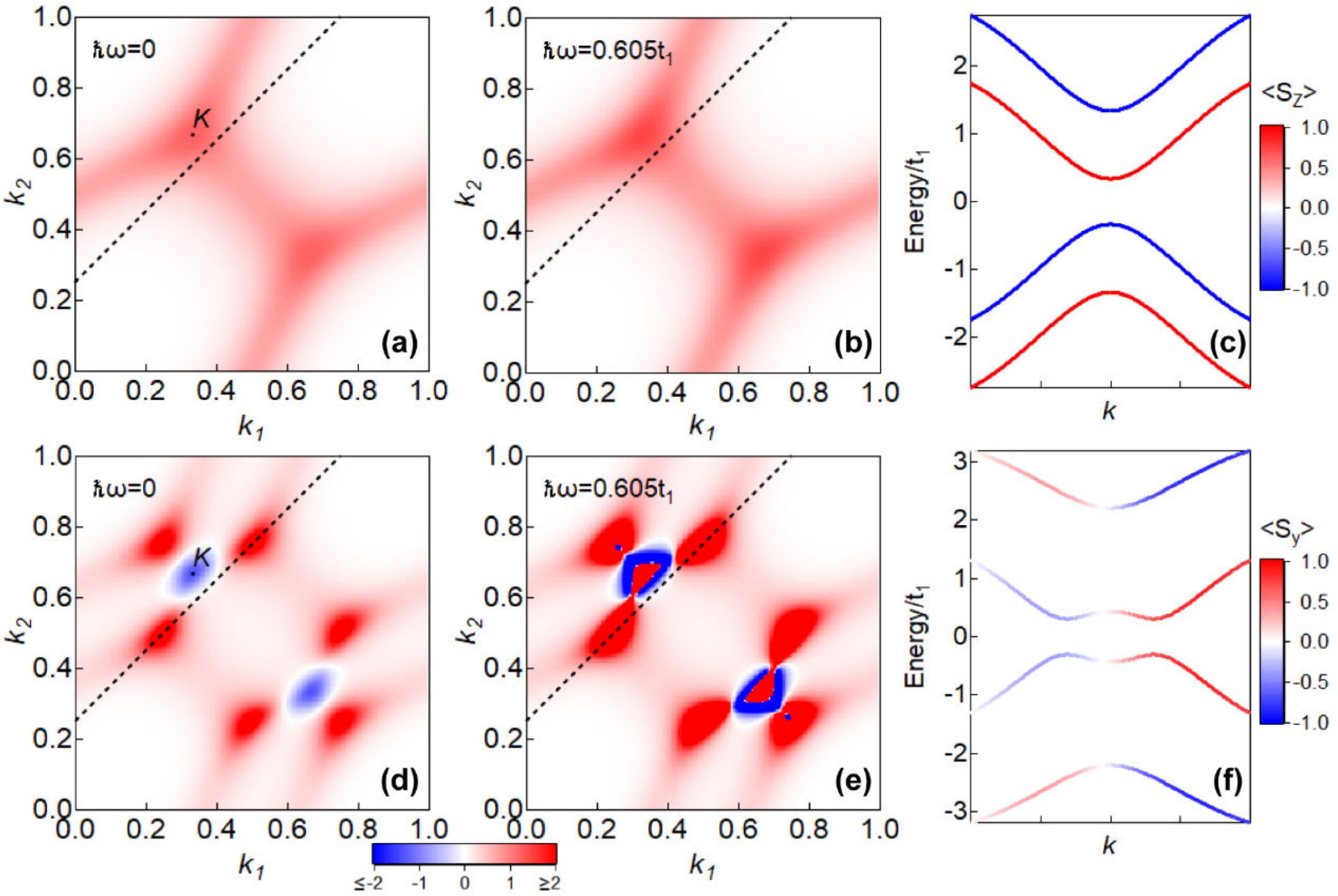

Fig. 6. (color online) Optical transition matrix elements and spin expectation values. (a) Berry curvature and (b) optical transition matrix element at $0.605t_1$ for $\lambda_R = 0$. (c) Corresponding bands along the dashed line in $k$-space from (a) and (b), with colors representing the expectation value of $S_z$. (d) Berry curvature and (e) optical transition matrix element at $0.605t_1$ for $\lambda_R = 0.6$. (f) Corresponding bands along the dashed line in $k$-space from (d) and (e), with colors representing the expectation value of $S_y$. $K$ point is labelled in (a) and (b). $k_1$ and $k_2$ are in units of the reciprocal lattice vector basis $\boldsymbol{b}_1$ and $\boldsymbol{b}_2$, respectively.

Since $\theta_F$ is essentially proportional to the real part of optical Hall conductivity $\sigma'_{xy}$, we further decomposed $\sigma'_{xy}$ into four kinds of channels: spin-conserving, spin-mixing, spin-flip and mixed spin-conserving/spin-flip channels. The numerator $[v_\mu(k)]_{nm}[v_\nu(k)]_{mn}$ in the Kubo-Greenwood formula can be separated into 16 terms based on different combinations of spin-up and spin-down components. Spin-conserving channels are explicitly given by $[v_\mu(k)]_{n\uparrow,m\uparrow}[v_\nu(k)]_{m\uparrow,n\uparrow}$ and $[v_\mu(k)]_{n\downarrow,m\downarrow}[v_\nu(k)]_{m\downarrow,n\downarrow}$, abbreviated as ↑↑↑↑ and ↓↓↓↓. In this notation, ↑ and ↓ denote the spin-up and spin-down components of the eigenstates $|\psi_n>$ and $|\psi_m>$.

Similarly, spin-mixing channels are expressed as ↑↑↓↓ and ↓↓↑↑; spin-flip channels as ↓↑↑↓, ↑↓↓↑, ↓↑↓↑ and ↑↓↑↓; and mixed spin-conserving/spin-flip channels as ↓↑↓↓, ↓↓↑↓, ↓↑↑↑, ↑↑↑↓, ↑↓↓↓, ↑↑↓↑, ↑↓↑↑ and ↓↓↓↑. Previous work by Misemer, which considered only weak intrinsic SOC in transition metals, concluded that only spin-conserving processes are important[10]. In contrast, a recent study on the effect of Rashba SOC and magnetic field on FR properties of $MoS_2$ found that spin-flip transitions lead to FR enhancement[11]. Our detailed decomposition analysis for the Haldane model reveals a different origin for the Rashba SOC-enhanced FR. We listed the four major contributions to the real part of optical Hall conductivity $\sigma'_{xy}$ for $\lambda_{FM} = -0.5$, $M = 0$ and $t_2$=0.15 in Fig. 7, with the corresponding spin components labeled on each subplot. Due to symmetry, the ↑↑↑↑ and ↓↓↓↓ channels yield identical contributions to $\sigma'_{xy}$; similarly, ↑↑↓↓ and ↓↓↑↑ are identical, and the mixed channels ↓↑↓↓, ↓↓↑↓, ↓↑↑↑, and ↑↑↑↓ also yield identical contributions. Note that Fig. 7 plots the individual contribution of each channel, not the sum. As shown in Fig 7 (a-c), pure spin-conserving, pure spin-mixing and mixed spin-conserving/spin-flip channels all share the same sign as the total optical Hall conductivity and thus contribute positively to the FR enhancement. Conversely, Fig. 7 (d) demonstrates that the pure spin-flip process yields an opposite contribution and is therefore detrimental to the FR angle enhancement.

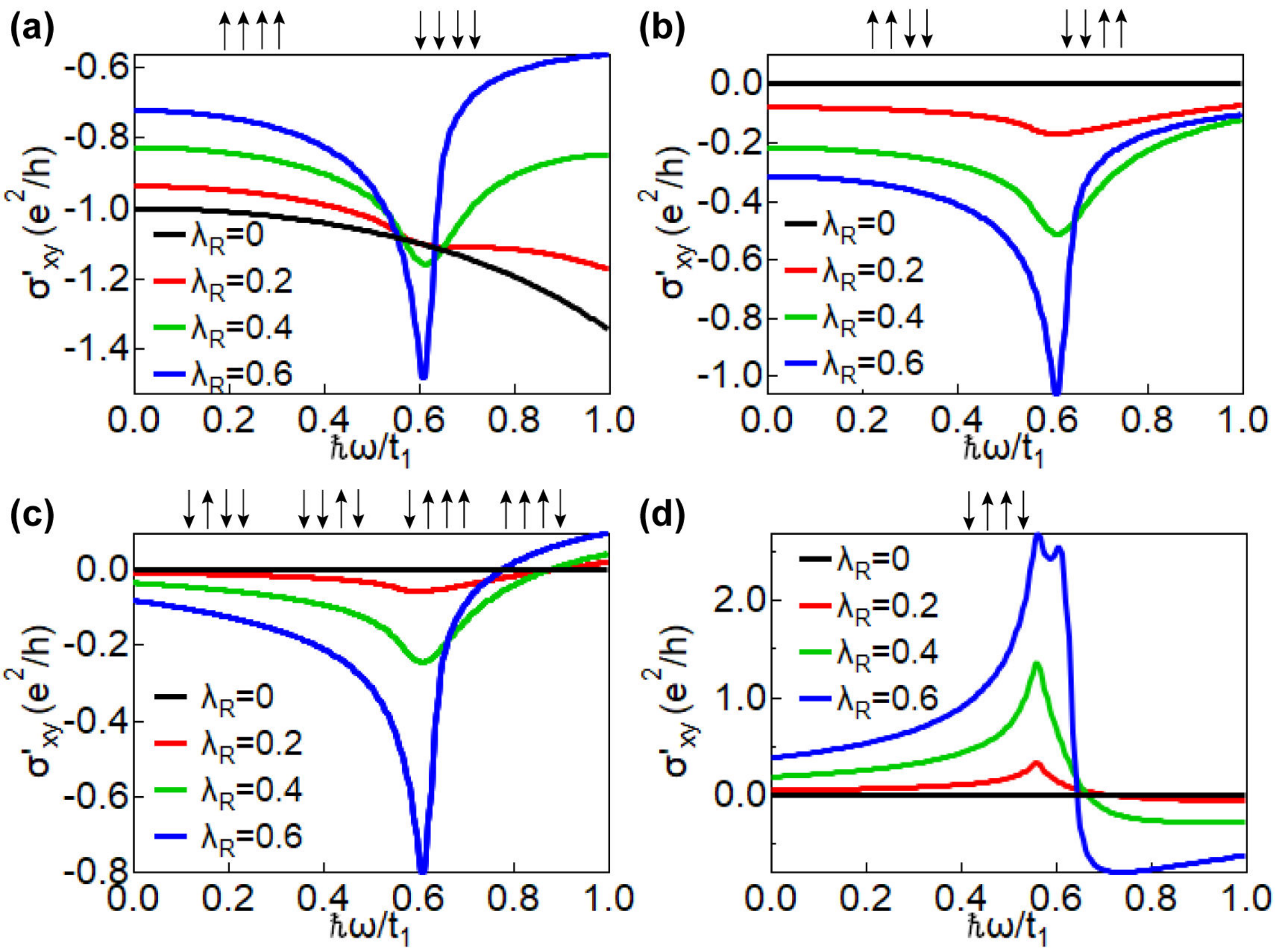

Fig. 7. (color online) Major contributions of different transition channels to the real part of optical Hall conductivity $\sigma'_{xy}$ for $\lambda_{FM} = -0.5$, $M = 0$ and $t_2$=0.15. (a) Pure spin-conserving contribution, (b) pure spin-mixing contribution, (c) mixed spin-conserving/spin-flip contribution, and (d) pure spin-flip contribution.

To explain the sign difference observed in Fig.7, we calculated the spin-channel resolved Berry curvature for the parameter set $\lambda_{FM} = -0.5$, $M = 0$, $t_2 = 0.15$ and $\lambda_R = 0.6$ as presented in Fig. 8. For the pure spin-flip channels ↓↑↑↓, the Berry curvature is predominantly concentrated near the $K$ and $K'$ points and is strictly negative throughout the Brillouin zone [Fig. 8(a)]. For the mixed spin-conserving/spin-flip channels, ↓↑↑↑ and ↓↑↓↓ yield identical Berry curvature distributions [Fig. 8(b)], while ↑↑↑↓ and ↓↓↑↓ share the same pattern [Fig. 8(c)]. In these mixed channels, the Berry curvature exhibits both positive and negative regions separated in momentum space, yet the positive contribution dominates, resulting in a net effect opposite to that of the pure spin-flip channel. For the pure spin-mixing channels, ↓↓↑↑ and ↑↑↓↓ produce identical positive Berry curvature over the whole momentum space [Fig. 8(d)].

Likewise, the pure spin-conserving channels ↓↓↓↓ and ↑↑↑↑ yield the same positive Berry curvature throughout the Brillouin zone [Fig. 8(e)].

To understand the physical origin of these signs, we recall that, from Fermi's golden rule, the difference in transition rates for left and right-handed circularly polarized lights is proportional to $Im(< f|v_x|i >< i|v_y|f >)$, where $|i >$ and $|f >$ denote the initial and final states, and $v_x$ and $v_y$ are x and y components of velocity operator. This quantity is the numerator in the Berry curvature definition [Eq. (4)]. Hence, the sign of the Berry curvature determines which circular polarization is predominantly absorbed: positive (negative) Berry curvature favors one handedness (the opposite). Consequently, the pure spin-flip channel predominantly absorbs the opposite circular polarization compared to the other dominant channels.

Finally, the Faraday rotation angle $\theta_F$ is related to the circular dichroism (absorption difference) through the Kramers–Kronig relation[31]. The absorption difference between left and right-handed circularly polarized lights is proportional to the imaginary part of $\sigma_{xy}$ [5] and for our parameters the FR peak always lies on the low-energy side of the absorption peak, as shown in Fig. S5. Therefore, a channel with positive Berry curvature (which absorbs one handedness light predominantly) yields a contribution to FR peak enhancement that is opposite in sign to that from a channel with negative Berry curvature. This accounts for the sign difference observed in Fig. 7.

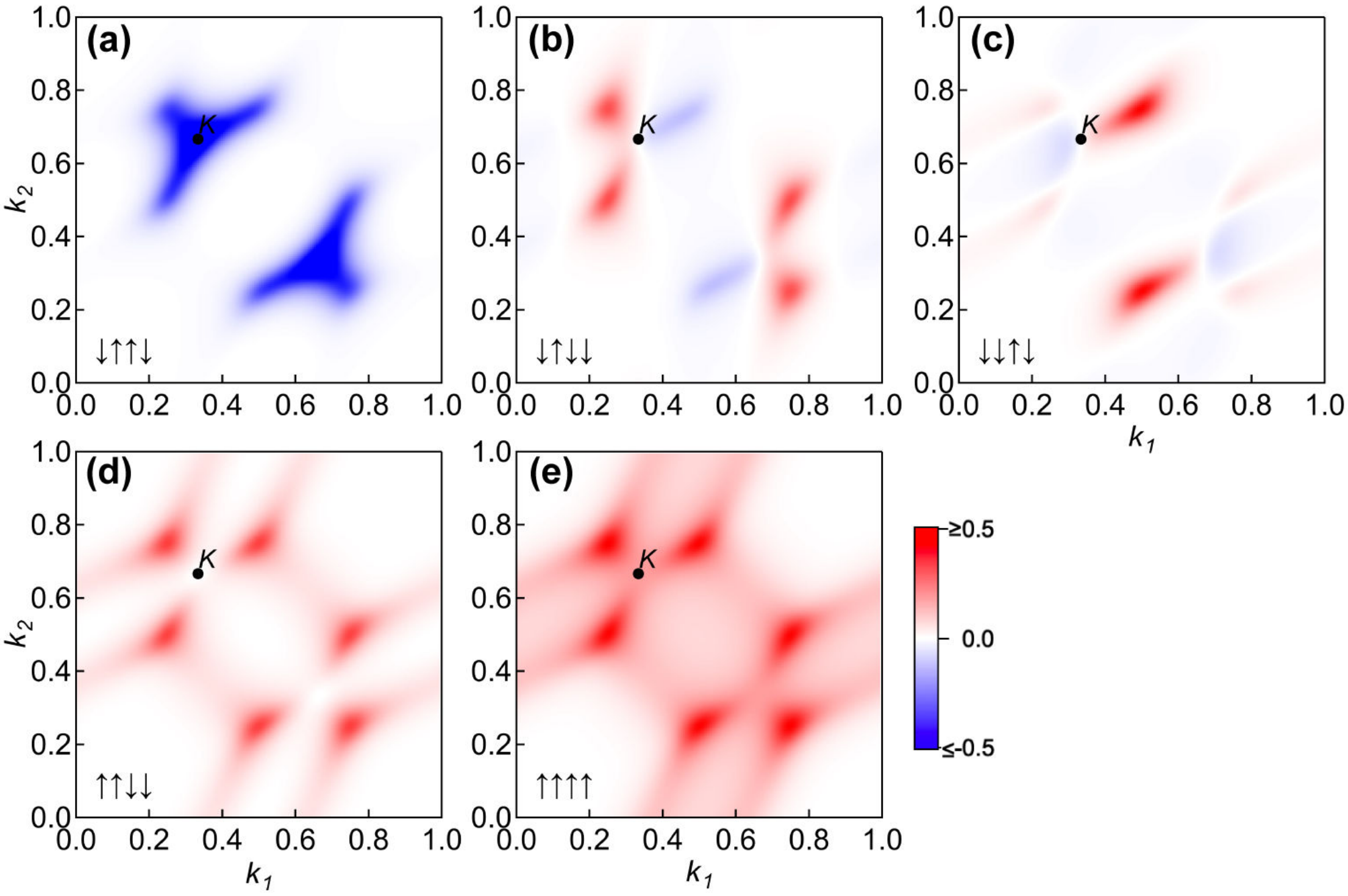

Fig. 8 Spin-channel resolved Berry curvature for $\lambda_{FM} = -0.5$, $M = 0$, $t_2 = 0.15$ and $\lambda_R = 0.6$. Spin channels are (a) ↓↑↑↓ (b) ↓↑↓↓ (c) ↓↓↑↓ (d) ↑↑↓↓ (e) ↑↑↑↑, where ↑ represents spin up and ↓ represents spin down. $k_1$ and $k_2$ are in units of the reciprocal lattice vector basis $\boldsymbol{b}_1$ and $\boldsymbol{b}_2$, respectively. $K$ point is labelled by the black dot.

### 3.3 Low-energy effective model

To corroborate our numerical calculations, we derived a low-energy effective Hamiltonian for the system from the tight-binding model[32,33] . Expanding up to linear terms in momentum $\boldsymbol{q}$ around the $K_\eta$ point, the effective Hamiltonian reads:

$$H_\eta^1 = -\frac{\sqrt{3}}{2} a t_1 (\eta q_x \tau_x + q_y \tau_y) + \frac{\sqrt{3}}{4} a \lambda_R [q_x (\tau_x \sigma_y + \eta \tau_y \sigma_x) + q_y (\tau_x \sigma_x - \eta \tau_y \sigma_y)] + \frac{3}{2} \lambda_R (\eta \tau_x \sigma_y - \tau_y \sigma_x) + 3\sqrt{3} \eta t_2 \tau_z + M \tau_z + \lambda_{FM} \sigma_z, \quad (11)$$

where we refer to the $K$ or $K'$ as the $K_\eta$ point with the valley index $\eta = \pm 1$ and $a$ is the lattice constant. The Hamiltonian is in the basis $\{\psi_{A\uparrow}, \psi_{A\downarrow}, \psi_{B\uparrow}, \psi_{B\downarrow}\}^t$. We found this linearized Hamiltonian $H_\eta^1$ exhibits significant deviations from the tight-binding results (not shown here). To achieve better quantitative agreement, we expanded the

Hamiltonian up to quadratic terms in momentum[34,35]. The quadratic correction terms are:

$$H_\eta^2 = \frac{a^2}{8} t_1 [(-{q_x}^2 + {q_y}^2)\tau_x + 2\eta q_x q_y \tau_y] - \frac{a^2}{8} \lambda_R [\frac{1}{2} ({q_x}^2 - {q_y}^2)(\eta \tau_x \sigma_y + \tau_y \sigma_x) + ({q_x}^2 + {q_y}^2)(\eta \tau_x \sigma_y - \tau_y \sigma_x) - q_x q_y (\eta \tau_x \sigma_x - \tau_y \sigma_y)] - \frac{3\sqrt{3} a^2}{4} \eta t_2 ({q_x}^2 + {q_y}^2)\tau_z, \quad (12)$$

We then computed the Berry curvature using the full effective Hamiltonian $H_\eta = H_\eta^1 + H_\eta^2$. Figure 9 compares the Berry curvature around $K$ point for $\lambda_R = 0$ and $\lambda_R = 0.6$, calculated with expansions up to linear [Figs. 9(a), 9(c)] and quadratic [Figs. 9(b), 9(d)] orders. The inclusion of quadratic terms significantly improves the agreement with Fig. 6(a) and 6(d). Finally, we compared the FR spectra calculated from the tight-binding model and the low-energy effective model (up to quadratic terms) for $\lambda_R = 0.4$, $t_2 = 0.15$, and four different sets of $M$ and $\lambda_{FM}$ values. As shown in Fig. 9(e), the results from the low-energy effective model (dots) match reasonably well with the tight-binding calculations (black curves), thus confirming the validity of our numerical approach.

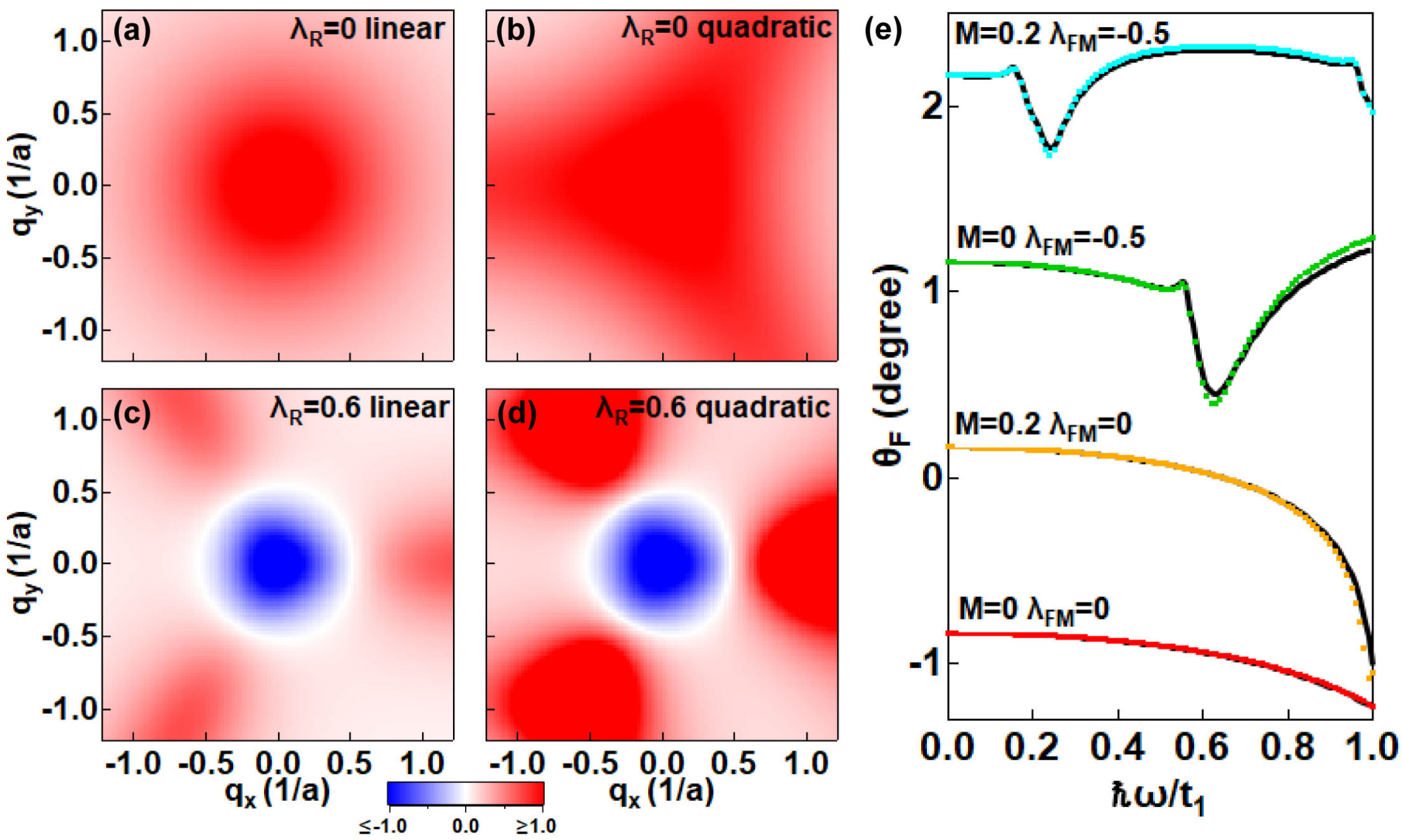

Fig. 9. (color online) Berry curvature for $\lambda_R = 0$ around $K$ point expanded up to (a) linear and (b) quadratic terms in momentum. Berry curvature for $\lambda_R = 0.6$ expanded

up to (c) linear and (d) quadratic terms. (e) FR curves for $\lambda_R = 0.4$ and $t_2 = 0.15$ for four sets of $M$ and $\lambda_{FM}$ values. Black curves represent tight-binding results and colored dots represent low-energy effective model results (up to quadratic terms). For clarity, the four datasets are vertically shifted by 1° with respect to each other.

### 3.4 Discussion on experimental feasibility

In our calculations, we set $t_1 = 1$ as the energy unit. The parameter ranges we explored— $t_2/t_1 \sim 0.01$ to 0.15, $\lambda_R/t_1 \sim 0$ to 0.6, $\lambda_{FM}/t_1 \sim -0.5$ to 0, and $M/t_1 \sim 0$ to 0.2—are chosen to cover the rich topological phase diagram of the extended Haldane model rather than to fit any specific material. Nevertheless, it is instructive to compare these values with existing platforms.

In graphene, intrinsic SOC is of the order of only 0.05 meV[36], and even the giant Rashba splitting induced by Au contact reaches only about 100 meV[37], corresponding to $\lambda_R/t_1 \sim 0.033$ with $t_1 \sim 3$ eV[38]. Proximity-induced exchange splitting in graphene can reach up to 30 meV[39], giving $\lambda_{FM}/t_1 \sim 0.01$. For silicene, germanene and stanene, the intrinsic Rashba SOC values are ~0.7 meV, 10.7 meV and 9.5 meV, respectively[40]. Still far below our typical $\lambda_R/t_1$ values. In transition metal dichalcogenide heterostructures, $\lambda_R/t_1$ can approach ~0.8 (e.g., in $MoS_2/Bi_2Te_3$)[41], but the exchange splitting achievable via proximity with magnetic insulators such as $CrBr_3$ is only about 14 meV[42], far less than $0.5t_1 \sim 500$ meV. Thus, although the experimental realization of a Haldane Chern insulator in AB-stacked MoTe2/WSe2 moiré bilayers was reported[43], in representative 2D solid-state materials the simultaneous realization of both strong Rashba SOC and strong exchange splitting remains a significant challenge.

Cold-atom systems offer a promising alternative, as the Haldane model has already been realized with ultracold fermions[44], and both Rashba SOC[45,46] and Zeeman fields[47] can be engineered for ultracold atoms. Our results may therefore serve as a useful reference for future experiments in these highly tunable platforms.

## 4. Conclusions

In conclusion, we have systematically investigated the effect of Rashba SOC on the FR properties of an extended Haldane model. We found that while FR can occur without Rashba SOC and exchange splitting, Rashba SOC plays a significant role in tuning the FR properties of the model. The FR in the Chern number $C = 2$ region exhibits particularly interesting behaviors. In the absence of exchange splitting, Rashba SOC can tune the peak positions of the FR spectra, which could serve as a fingerprint of the Rashba SOC strength. In the presence of exchange splitting, Rashba SOC can induce the emergence of flat spectra over a broad frequency range while simultaneously enhancing the FR peak values. Through calculations of the Berry curvature and finite-frequency optical transition matrix elements, we inferred that this enhancement arises from the opening of originally forbidden transition channels at finite Rashba SOC. Decomposition of the optical Hall conductivity reveals that pure spin-conserving, pure spin-mixing and mixed spin-conserving/spin-flip channels contribute significantly to the FR enhancement, whereas the pure spin-flip process is detrimental. This opposite contribution stems from the predominant absorption of opposite circularly polarized light. Results from a low-energy effective model expanded up to quadratic terms in momentum are in good agreement with those from the tight-binding model, corroborating our conclusions. Our study highlights the potential of Rashba SOC engineering as a powerful strategy for designing future magneto-optical devices.

Before ending, we want to mention that our work can be generalized to the study of FR properties in other model systems, such as the time reversal symmetry broken Kane-Mele model merged with the Haldane model[48] , multilayer systems[49] , 2D altermagnets[50] , etc.

## Acknowledgements

Yuan Fang thanks Zhao Liu and Siqi Wu for helpful discussions. Xiaopu Zhang and Yuan Fang acknowledge the funding support from National Natural Science Foundation of China (Grant No. 52573361).